\documentclass[iop]{emulateapj}

\usepackage{amsmath}
\usepackage{amssymb}
\usepackage{mathbbol}
\usepackage{dsfont}

\shorttitle{Persistence of two small-scale $\Lambda$CDM problems}
\shortauthors{Pawlowski et al.}

\begin{document}

\title{On the persistence of two small-scale problems in $\Lambda$CDM}

\author{Marcel S. Pawlowski\altaffilmark{1}}
\author{Benoit Famaey\altaffilmark{2}}
\author{David Merritt\altaffilmark{3}}
\author{Pavel Kroupa\altaffilmark{4}}

\affil{$^1$Department of Astronomy, Case Western Reserve University,\\
10900 Euclid Avenue, Cleveland, OH, 44106, USA\\
$^2$Observatoire astronomique de Strasbourg, Universit\'e de Strasbourg,\\
CNRS, UMR 7550, 11 rue de l'Universit\'e, F-67000 Strasbourg, France\\
$^3$School of Physics and Astronomy and Center for Computational Relativity and Gravitation,\\
Rochester Institute of Technology, 84 Lomb Memorial Drive, Rochester, NY 14623, USA\\
$^4$Helmholtz-Institut f\"ur Strahlen- und Kernphysik, Rheinische Friedrich-Wilhelms-Universit\"at Bonn,\\
Nussallee 14-16, D-53115 Bonn, Germany}

\email{marcel.pawlowski@case.edu}

\begin{abstract}
We investigate the degree to which the inclusion of baryonic physics can overcome two long-standing 
problems of the standard cosmological model on galaxy scales: (i) the problem of satellite planes around 
Local Group galaxies, and (ii) the ``too big to fail'' problem.
By comparing dissipational and dissipationless simulations, we find no indication that the addition of 
baryonic physics results in more flattened satellite distributions around Milky-Way-like systems. 
Recent claims to the contrary are shown to derive in part from a non-standard metric for the degree
of flattening, which ignores the satellites' radial positions.
If the full 3D positions of the satellite galaxies are considered, none of the simulations we analyse reproduce the 
observed flattening nor the observed degree of kinematic coherence of the Milky Way satellite system.
Our results are consistent with the expectation that baryonic physics should have little or no influence on 
the structure of satellite systems on scales of hundreds of kiloparsecs.
Claims that the ``too big to fail'' problem can be resolved by the addition of baryonic physics are also
shown to be problematic.
\end{abstract}


\section{Introduction}
The $\Lambda$CDM model of cosmology has long been known to be plagued by numerous `small-scale-problems'. These include the much larger predicted number of sub-halos compared to the number of observed satellite galaxies (missing satellites problem, \citealt{Klypin1999,Moore1999}), the discrepancy between a predicted density peak of dark matter at the centers of halos and observations indicating a constant-density core (core-cusp problem, \citealt{DubinskiCarlberg1991,WalkerPenarrubia2011}), the apparent over-abundance of backsplash galaxies in the Local Group \citep{PawlowskiMcGaugh2014a}, and many other tensions between $\Lambda$CDM predictions and observations (summarized for example by \citealt{Kroupa2012,FamaeyMcGaugh2013,McGaugh2014,Weinberg2013} and \citealt{WalkerLoeb2014}). Two problems which have emerged recently and are gaining increasing attention are the Too-Big-To-Fail (TBTF) and the Satellite-Planes problems.

The TBTF problem is concerned with the internal dynamics of dwarf galaxies. When comparing the central masses of MW dwarf Spheroidal (dSph) satellites deduced from their kinematics with those of dark matter sub-halos in simulations, \citet{BoylanKolchin2011,BoylanKolchin2012} found that simulations of MW equivalents each contain $\approx10$\ (actually 5 to 40) denser sub-halos than those compatible with the most-luminous observed dSphs. This can be related to Fig.~2 of Kroupa et al.~(2010) where it was shown that the observationally deduced DM halo masses of the MW satellites show a significant overabundance of $M_{0.3kpc} \simeq 10^7 M_\odot$ halos and a lack of both less and more massive values compared to the theoretically predicted distribution for luminous sub-halos. This would indicate that the most-massive sub-halos do not host the most-luminous dSphs but remain virtually dark, while the most-luminous dSphs live in sub-halos of only intermediate mass. This raises the question of what prevented the massive sub-halos from forming galaxies. The TBTF problem has been identified not only among the MW satellite galaxies, but also for the M31 satellite galaxies \citep{Tollerud2014}, within the Local Group \citep{Kirby2014,GarrisonKimmel2014b} and also appears to be present for field galaxies \citep{Papastergis2015}.

The TBTF problem is more difficult to resolve than the missing satellites problem since it requires either that luminous galaxies do not form in the sub-halos with the largest central dark matter density, or that one introduces a process which reduces the central dark matter density of these most massive sub-halos to values consistent with the observed velocities in dwarf galaxies. One suggested solution to this problem has been a `light' MW. If the MW halo is less massive than generally assumed (at least $\lesssim 8 \times 10^{11} M_\sun$\ instead of 1 to $2 \times 10^{12} M_\sun$) this translates into a lower number of massive sub-halos, thus alleviating the TBTF problem \citep{BoylanKolchin2012,Wang2012,VeraCiro2013}. However, in general such a low MW mass is disfavoured, for example by the Local Group timing argument \citep{LiWhite2008}, the analysis of the positions, line-of-sight velocities and proper motions of the MW satellite population \citep{Watkins2010}, the Galactic escape speed as a function of radius \citep{Piffl2014}, and the modelling of stellar streams in the MW halo \citep{Kuepper2015}. Furthermore, that the TBTF problem is also present for the M31 satellites \citep{Tollerud2014}, for more distant dwarfs in the Local Group \citep{Kirby2014,GarrisonKimmel2014b} and possibly even beyond \citep{Papastergis2015} is a strong argument against purely local or environment-dependent solutions.

But not only are the internal properties of dwarf galaxies problematic for $\Lambda$CDM. The overall spatial distribution of satellite galaxies also appears to be at odds with cosmological predictions. \citet{KunkelDemers1976} and \citet{LyndenBell1976} were the first to realize that the then-known satellites of the MW are anisotropically distributed and align in a common plane with the Magellanic Stream. \citet{Kroupa2005} first argued that this flattened distribution of the 11 brightest (`classical') MW satellites is problematic for the cosmological model. 
Not only are these 11 classical satellites situated in this flattened structure, but also the later discovered faint and ultra-faint satellites define the same structure \citep{Metz2009b,Kroupa2010}. Furthermore, \citet{Pawlowski2012a} have shown that this `Vast Polar Structure' (VPOS) also consists of globular clusters categorized as young halo objects (which are likely associated with satellite galaxies), and that about half of the stellar and gaseous streams in the MW halo align with the satellite plane. The recently discovered MW satellites in the southern hemisphere have also been found to align with the VPOS \citep{Pawlowski2015}. Finally, the potentially most important piece of information is provided by the proper motions measured for the 11 brightest satellites. These reveal that nine of the satellites are consistent with orbiting within the VPOS, eight of these even co-orbit in the same direction \citep{Metz2009,PawlowskiKroupa2013}. 

A similar plane of satellite galaxies consisting of about half the M31 satellites has been discovered by \citet{Ibata2013}. It is seen edge-on from the MW. \citet{Ibata2013} found that 13 out of 15 satellite in this Great Plane of Andromeda (GPoA) show coherent line-of-sight velocities, indicating that they might co-orbit M31, analogous to the satellites in the VPOS. In addition to the satellite galaxies, the isolated dwarf galaxies in the Local Group are also confined to two very narrow and highly symmetric planes \citep{Pawlowski2013a}. Similarly, the dwarf galaxy members of the nearby NGC 3109 association are confined to a very narrow, linear distribution \citep{Bellazzini2013, PawlowskiMcGaugh2014a}. Coherent alignments of satellite galaxies and streams around more distant hosts have also been discovered in recent years \citep[e.g. summarized in][]{PawlowskiKroupa2014}. Most recently \citet{Tully2015} have reported two narrow dwarf galaxy planes around Centaurus A, and \citet{Mueller2015} reported the discovery of 16 dwarf galaxy candidates distributed anisotropically around around M83. \citet{Ibata2014b,Ibata2014c} have performed a systematic study of diametrically opposite satellite galaxies around hosts within the Sloan Digital Sky Survey \citep[SDSS][]{York2000} and found that they preferentially display anti-correlated line-of-sight velocities, which is consistent with approximately 60 per cent of all satellite galaxies being situated in thin, co-orbiting planes. This interpretation has been challenged by \citet{Phillips2015}, who performed a similar analysis of SDSS satellite pairs and find a signal consistent with only 30\,per cent of all satellites to be in co-orbiting planes. Another related analysis was performed by \citet{Cautun2015}, who suggest that the observed signal is sensitive to selection effects. Given this situation, one will have to wait for further observational data, increasing the number of opposite satellite pairs with known line-of-sight velocities, to put tighter constraints on the fraction of co-orbiting satellites in the Universe. But the existence of satellite-plane configurations in the Local Group is not subject to debate.

The observed narrowness of the satellite galaxy planes of the Local Group and their coherent kinematic signature have been found to be extremely unlikely among satellite galaxies drawn from a variety of cosmological simulations. \citet{Metz2007} found the spatial distribution of the then known MW satellites to be inconsistent with an isotropic or prolate dark matter substructure distribution at a 99.5 per cent level. \citet{Pawlowski2012b} determined that the correlation and polar orientation of satellite orbits seen among the eight satellites for which proper motion measurements where available at that time was very rare (0.4 to 2 per cent) among dark matter sub-halos drawn from from the Via Lactea \citep{Diemand2007,Diemand2008} and Aquarius simulations \citep{Springel2008}. A later re-analysis \citep{PawlowskiKroupa2013} used improved proper motion information and included three additional satellites for which proper motions were determined in the meantime. They found that the observed kinematic coherence of the 11 brightest MW satellites alone is present in only 0.6 per cent of satellite systems found in the Via Lactea and Aquarius simulations, and is also very unlikely when compared to the $\Lambda$CDM predictions of \citet{Libeskind2009} and \citet{Deason2011}. \citet{Ibata2014a} determined that finding a GPoA-like structure with a similar kinematic signature in the Millennium-II simulation has a chance of only 0.04 per cent. This result was confirmed by \citet{Pawlowski2014}, who also demonstrated that a satellite structure as flattened and coherently orbiting as the VPOS is found in only 0.06 per cent of all cases in this simulation. This is a conservative upper limit since the analysis used only the spatial and kinematic alignment of the 11 brightest satellites. Most recently, \citet{Gillet2015} were unable to find a GPoA equivalent in their analysis of a simulation of a Local Group equivalent from the CLUES project. Their results also show that simulated satellites which can be assigned to a planar distribution for a given snapshot of a simulation display large velocities perpendicular to that plane. While not testable for the M31 case, this is inconsistent with the observed alignment of the orbital directions of the MW satellites in the VPOS. A similar analysis by \citet{PawlowskiMcGaugh2014b} used the `Exploring the Local Volume in Simulations' (ELVIS) suite of simulations by \citet{GarrisonKimmel2014a} and found that only one out of 4800 realizations (0.02 per cent) was able to reproduce the observed flattening and orbital correlation of the 11 brightest MW satellites. They also conclude that host galaxies in Local Group environments are as unlikely to harbour satellite galaxy planes as isolated hosts. In contrast to these analyses, claims of consistency of the observed satellite galaxy planes have frequently been shown to ignore aspects such as the kinematic coherence of the observed structure, to be based on unrealistic assumptions, or to not properly account for survey selection effects \citep[e.g.][]{Metz2009,Pawlowski2012b,Ibata2014a, Pawlowski2014}. 

The small-scale problems are often not seen as immediate failures of the $\Lambda$CDM model, but rather as possibly caused by flaws in constructing $\Lambda$CDM predictions from non-dissipative 'dark-matter-only' simulations (but see e.g. \citealt{Kroupa2012,Kroupa2015}). On scales of galaxies, the effects of baryons such as cosmic reionization, star formation and feedback from stars, supernovea or AGN become important \citep{Navarro1996,Bullock2000,Mashchenko2008,BovillRicotti2009,Governato2010,Sawala2010,Pontzen2012}. Consequently there are now numerous attempts to include baryonic effects in cosmological simulations by modelling the hydrodynamics of the gas, as well as star formation and feedback processes on a sub-grid level \citep[e.g.][]{Libeskind2010,Vogelsberger2014}. Unfortunately these models often suffer from unphysical modelling of the baryonic processes \citep[see for example the discussions in][]{Kroupa2015, Schaye2015}.

While the effect of baryons can be substantial for $\Lambda$CDM predictions and cause a re-evaluation of some of the problems (such as the missing satellites problem, \citealt{Koposov2009}; or the core-cusp problem, \citealt{Pontzen2012}; but see \citealt{Penarrubia2012} why solutions to these two problems might be mutually exclusive), others are more independent of baryonic effects. This is particularly true for the satellite plane problem: the distribution of satellite dwarf galaxies and their kinematic coherence on scales of hundreds of kpc -- where the gravitational effect of the dark matter is dominant -- can be safely assumed to not be affected by the internal physics of individual dwarf galaxies.

In this respect it can be seen as surprising that \citet[][S14 hereafter]{Sawala2014} have claimed that a set of hydrodynamic cosmological simulations they analyze resolves not only the missing satellites and TBTF problems, but also the satellite plane problem. Their work is based on the `Evolution and Assembly of GaLaxies and their Environments' (EAGLE) simulations by the Virgo Consortium \citep{Schaye2015}. They analyze a set of hydrodynamical cosmological zoom-simulations leading to the formation of Local Group-like galaxy pairs, which model a number of baryonic physics effects (see \citealt{Schaye2015} and the supplementary information of S14 for more details). 

In Sect. \ref{sect:planes} we investigate whether the measure of flattening of a satellite system from the satellites' projected positions \citep[][S14]{Starkenburg2013} provides a reliable test of similarity to the observed situation (Sect. \ref{sect:positions}), whether the distribution of orbital poles of simulated satellite systems shown by S14 is similarly correlated as those of the observed MW satellites (Sect. \ref{fig:poles}) and whether the range of flattening of satellite systems found in hydrodynamic cosmological simulations are indeed different to purely collision-less 'dark-matter-only' simulations of Local Group equivalents (Sect. \ref{sect:dmonly}). We will then discuss the persistence of the TBTF problem in Sect. \ref{sect:TBTF}. Our results will be summarized and discussed in Sect. \ref{sect:conclusion}.


\section{Satellite Plane Problem}
\label{sect:planes}

The satellite plane problem for the eleven classical MW satellite galaxies\footnote{Confining the analysis to these 11 brightest satellites avoids the issue that less luminous satellites were discovered in surveys with uneven sky coverage and is in line with the analysis by S14.} 
can be summarized as follows: The (three-dimensional) positions of the MW satellite galaxies are distributed such that they all lie close to one common plane which has a root-mean-square (rms) height of 19.6\,kpc and a rms minor-to-major axis ratio of $c/a = 0.182$. Most of these satellites have orbital planes which are closely aligned with the plane defined by their position and they also share the same orbital direction (they co-orbit). This is indicated by the close clustering of 8 out of 11 orbital poles (directions of angular momentum) close to one normal vector describing the orientation of the best-fitting plane. In addition, one of the remaining three of the 11 orbital poles is directed along the opposite pole (i.e. retrograde relative to the 8 others).

Therefore, the defining characteristics of the satellite plane problem are that:
\begin{itemize}
\item the satellites are distributed in a highly flattened, \textit{planar} structure in three-dimensional space,
\item the \textit{majority} of the satellites \textit{co-orbit} in the same sense,
\item and these satellites orbit \textit{within} the plane, indicating that the plane is not just a transient alignment.
\end{itemize}

\subsection{Positions}
\label{sect:positions}

The surprising degree of flattening of the MW satellite system has commonly been described in either absolute terms, as the rms (minor-axis) height of e.g. the 11 classical MW satellites ($r_{\mathrm{per}} = 19.6$\,kpc, e.g. \citealt{Kroupa2005, Zentner2005, Metz2007, Metz2009, Kroupa2010, Wang2012, Pawlowski2013a, Pawlowski2014}), or as a relative flattening of the distribution's minor-to-major axis ratio ($(c/a)_{\mathrm{std}}^{\mathrm{MW}} = 0.18$, e.g. \citealt{Metz2007, Libeskind2005, Libeskind2009}\footnote{Note that accoding to \citet{Wang2013} they had an error in their code calculating $c/a$.}, \citealt{Deason2011, Wang2013, Pawlowski2013a, Pawlowski2014, PawlowskiMcGaugh2014b}). Commonly the major and minor axis directions are determined from the eigenvectors of the tensor of inertia (ToI) defined by the non-weighted satellite positions. 

\citet{Starkenburg2013} and S14 chose to define flattening in a different and idiosyncratic way.
They characterize the spatial anisotropy of a satellite system using the ratio of eigenvalues of the 
\textit{reduced} ToI \citep{BailinSteinmetz2005}. For the 11 classical MW satellites, this results in\footnote{Note that we identify $c$\ and $a$\ with the root-mean-square heights of the satellite distribution along the minor and major axes, while S14 identify them with the eigenvalues of the reduced ToI, in which case the axis ratio would be $\sqrt{(c'/a')_{\mathrm{Eigenvalues}}}$.} 
$(c/a)_{\mathrm{red}}^{\mathrm{MW}} = 0.36$. 
Their approach is equivalent to measuring the axis ratios of the satellite distribution \textit{after the satellites have been projected onto a unit sphere around the host galaxy's center}. 
Such a measure does not characterize a planar flattening of the satellite system, but only the preference of satellites to lie close to one common great circle, or to cluster about two opposed directions on the sky. 

No justification was provided by these authors for their non-standard way of characterizing the 
``flattening'' of a satellite system. 
S14  refer to \citet{BailinSteinmetz2005}. That study used the reduced ToI to measure and compare the principal axes of the dark matter particle mass distribution among six concentric shells within dark matter halos. The radial distance normalization was introduced so that substructures in the outer part of a shell would not dominate the ToI. These arguments do not apply to the satellite plane problem, which is not concerned with the principal axes of the \textit{mass} distribution but merely with the distribution of satellite positions (each satellite carries the same weight regardless of its mass); substructure is not an issue in the satellite distribution (which consists only of $\sim 10^1$  objects), and no comparison between different radial shells is intended (only one axis ratio of the whole satellite system is desired).

One potential benefit of the reduced ToI is that it might be less sensitive to outliers at large distances. For the observed MW satellites, the most distant satellite Leo I is possibly not associated with the VPOS, because its proper motion results in a most-likely orbital pole not aligned with the satellite plane normal (see Sect. \ref{sect:poles}). Furthermore, Leo I is rapidly receding from the MW, such that -- depending on the potential of the MW -- it might not even be a bound satellite of the MW \citep{BoylanKolchin2013}\footnote{Interestingly, as discussed in \citet{PawlowskiMcGaugh2014a}, Leo I's most-likely velocity vector is consistent with it moving into and along the ''Great Northern Plane'', the narrow planar arrangement consisting of all non-satellite dwarf galaxies in the northern hemisphere of the MW, which also seems to hosts a puzzling over-abundance of such backsplash galaxies \citep{PawlowskiMcGaugh2014a}.}. To test the sensitivity of the standard and reduced ToI on this most-distant satellite, we have repeated the ToI fits using only the 10 classical MW satellites excluding Leo I. As expected, the axis ratio of the reduced ToI fit is not strongly affected ($(c/a)_{\mathrm{std}}^{\mathrm{no LeoI}} = 0.37$, compared to 0.36 including Leo I), but neither is that of the standard ToI ($(c/a)_{\mathrm{std}}^{\mathrm{no LeoI}} = 0.20$, compared to 0.18 including Leo I). The orientation of the satellite plane derived using the reduced ToI changes by less than one degree, and that derived using the standard ToI by $6.4^{\circ}$. For the latter, excluding Leo I results in a slightly better alignment with the average orbital poles in Fig. \ref{fig:poles}.

However, the standard ToI fit is less affected by the satellites closest to the MW. For a non-infinitesimally thin, planar structure these nearby satellites are expected to be the most offset from the overall distribution in their angular positions. Furthermore, they reside in a region in which processes which have the potential to change their orbit, such as precession and satellite-satellite encounters, are more important. Thus, while maybe less sensitive to distant outliers, the reduced ToI is \textit{more} sensitive to outliers at close distances. In the case of the MW this can be demonstrated using Sagittarius, which has a Galactocentric distance of only 18\,kpc but is orbiting perpendicular to the VPOS. Applying the two ToI methods to the 10 classical satellites except Sagittarius results in almost the same axis ratios for the standard ToI ($(c/a)_{\mathrm{std}}^{\mathrm{no Sag}} = 0.17$, compared to 0.18 including Sagittarius), but in largely different parameters for the reduced ToI fit ($(c/a)_{\mathrm{red}}^{\mathrm{no Sag}} = 0.25$, compared to 0.36 including Sagittarius). Similarly, the best-fit plane orientation is essentially unchanged for the standard ToI (the directions differ by only $2.0^{\circ}$), but for the reduced ToI the orientation is changed by $9.5^{\circ}$\ if Sagittarius is excluded from the fit. Thus, compared to the standard ToI the reduced ToI trades a somewhat decreased sensitivity to distant outliers for a higher sensitivity to nearby outliers.

In the following, we will show that the use of the reduced ToI ignores available information and is a less robust and less meaningful test of the flattening of a satellite system.

If the position vector of satellite $i$ is $\mathbf{r}_i = (r_i^\mathrm{x}, r_i^\mathrm{y}, r_i^\mathrm{y})$, the standard ToI is
\begin{equation}
\begin{split}
I_{\mathrm{std}}(\mathbf{r}) & = \sum^{11}_{i=1} \left(\mathbf{r}_i^2 \cdot \mathbb{1}\right) - \left(\mathbf{r}_i \mathbf{r}_i^{\mathrm{T}}\right) \\
& = \sum^{11}_{i=1}
\begin{pmatrix}
\left(r_i^\mathrm{y}\right)^2 + \left(r_i^\mathrm{z}\right)^2 & -r_i^\mathrm{x} r_i^\mathrm{y} & -r_i^\mathrm{x} r_i^\mathrm{z}\\
-r_i^\mathrm{x} r_i^\mathrm{y} & \left(r_i^\mathrm{x}\right)^2 + \left(r_i^\mathrm{z}\right)^2 & -r_i^\mathrm{y} r_i^\mathrm{z}\\
-r_i^\mathrm{x} r_i^\mathrm{z} & -r_i^\mathrm{y} r_i^\mathrm{z} & \left(r_i^\mathrm{x}\right)^2 + \left(r_i^\mathrm{y}\right)^2
\end{pmatrix}
\end{split}
\end{equation}
where $\mathbf{r}_i^{\mathrm{T}}$ is the transposed position vector $\mathbf{r}_i$.
The elements of the reduced ToI are defined as \citep{BailinSteinmetz2005}:
\begin{equation}
I_{\mathrm{red}}^{\alpha,\beta}(\mathbf{r}) = \sum^{11}_{i=1} \frac{r_i^\alpha r_i^\beta}{r_i^2},
\end{equation}
so the total reduced ToI is
\begin{equation}
I_{\mathrm{red}}(\mathbf{r}) = \sum^{11}_{i=1}
\begin{pmatrix}
\left(r_i^\mathrm{x}\right)^2 & r_i^\mathrm{x} r_i^\mathrm{y} & r_i^\mathrm{x} r_i^\mathrm{z}\\
r_i^\mathrm{x} r_i^\mathrm{y} & \left(r_i^\mathrm{y}\right)^2 & r_i^\mathrm{y} r_i^\mathrm{z}\\
r_i^\mathrm{x} r_i^\mathrm{z} & r_i^\mathrm{y} r_i^\mathrm{z} & \left(r_i^\mathrm{z}\right)^2
\end{pmatrix} / \mathbf{r}_i^2.
\end{equation}
Therefore, if each position vector $\mathbf{r}_i$\ is normalized to $\mathbf{r'}_i = \frac{\mathbf{r}_i}{\left|\mathbf{r}_i\right|}$, then $I_{\mathrm{red}}(\mathbf{r'}) = I_{\mathrm{red}}(\mathbf{r})$ and
\begin{equation}
\begin{split}
I_{\mathrm{std}}(\mathbf{r'}) + I_{\mathrm{red}}(\mathbf{r'}) & = \sum_{i=1}^{11} \left( \left(r_i^\mathrm{x}\right)^2 + \left(r_i^\mathrm{y}\right)^2 + \left(r_i^\mathrm{z}\right)^2 \right) \mathbb{1} \\
& = \sum_{i=1}^{11} \mathbb{1}.
\end{split}
\end{equation}
We can see that the standard ToI $I_{\mathrm{std}}$\ constructed using the normalized satellite positions is equivalent to the negative of the reduced ToI, $I_{\mathrm{red}}$, except for an additive factor proportional to the unit matrix $\mathbb{1}$. This has the important consequence that eigenvectors of $I_{\mathrm{std}}$\ constructed from the normalized satellite positions are identical to those of $I_{\mathrm{red}}$\ (every vector is an eigenvector to the unit matrix). Either ToI will therefore give the same axes of most and least flattening of a satellite system with normalized positions. Due to the negative sign in this relation, the eigenvector corresponding to the shortest axis of the satellite distribution belongs to the smallest eigenvalue of $I_{\mathrm{red}}$, but to the largest eigenvalue of $I_{\mathrm{std}}$. In the following we will use both the reduced ToI and the more commonly employed standard ToI to compute the respective axis ratios $(c/a)_{\mathrm{red}}$\ and $(c/a)_{\mathrm{std}}$\ of satellite systems.

In spite of this connection, however, the reduced ToI $I_{\mathrm{red}}$\ contains less information than the
standard ToI, because only two angular coordinates of the satellites are used, not the full positions.
It can thus be expected that using $(c/a)_{\mathrm{red}}$\ as a metric for flattening of satellite systems 
will be less robust than a test based on $(c/a)_{\mathrm{std}}$. We now demonstrate this.
In what follows, we focus on the 11 brightest, classical MW satellites, whose positions we take from the compilation of \citet{McConnachie2012}.

\subsubsection{Random satellite positions from an isotropic distribution}

\begin{figure*}
   \centering
   \includegraphics[width=80mm]{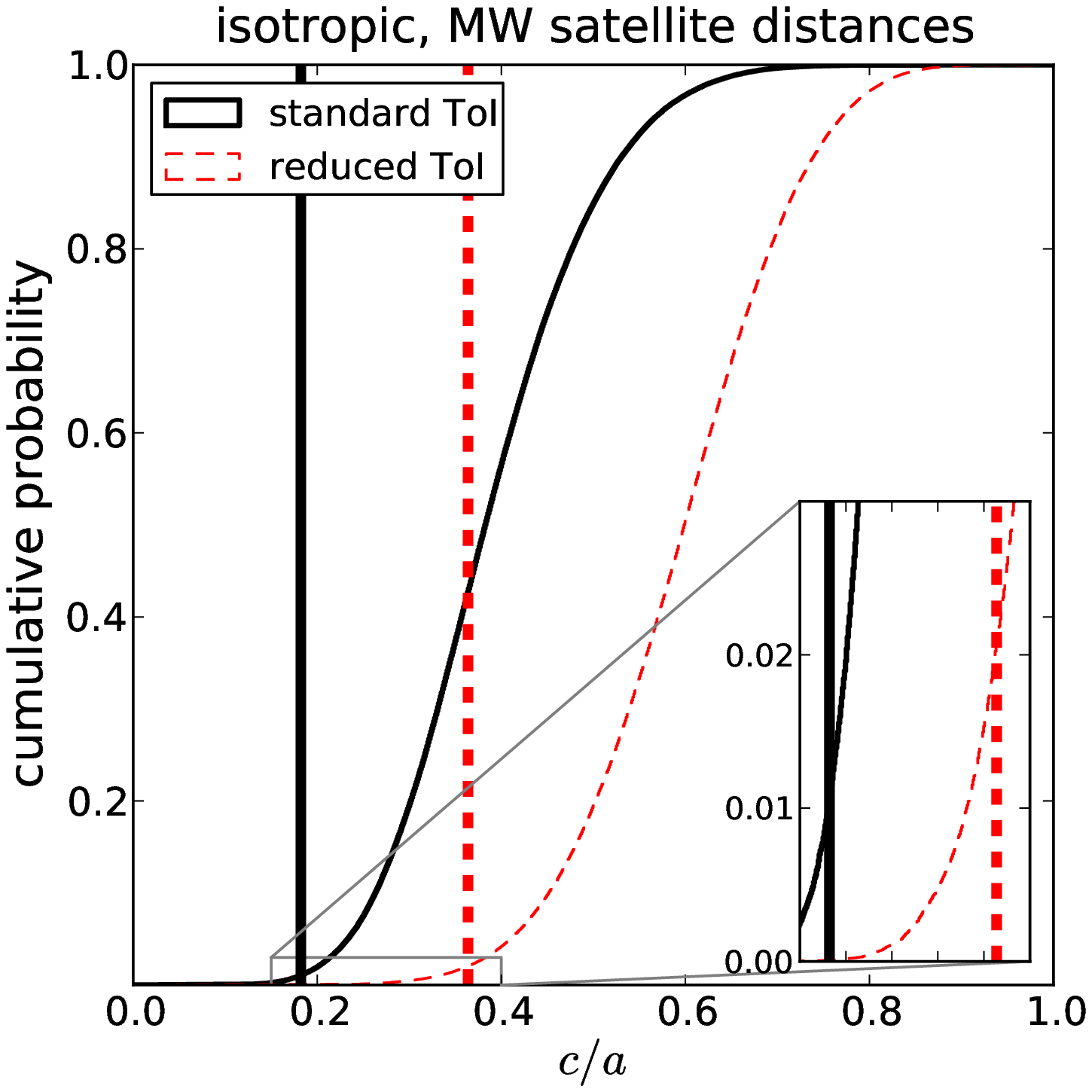}
   \includegraphics[width=80mm]{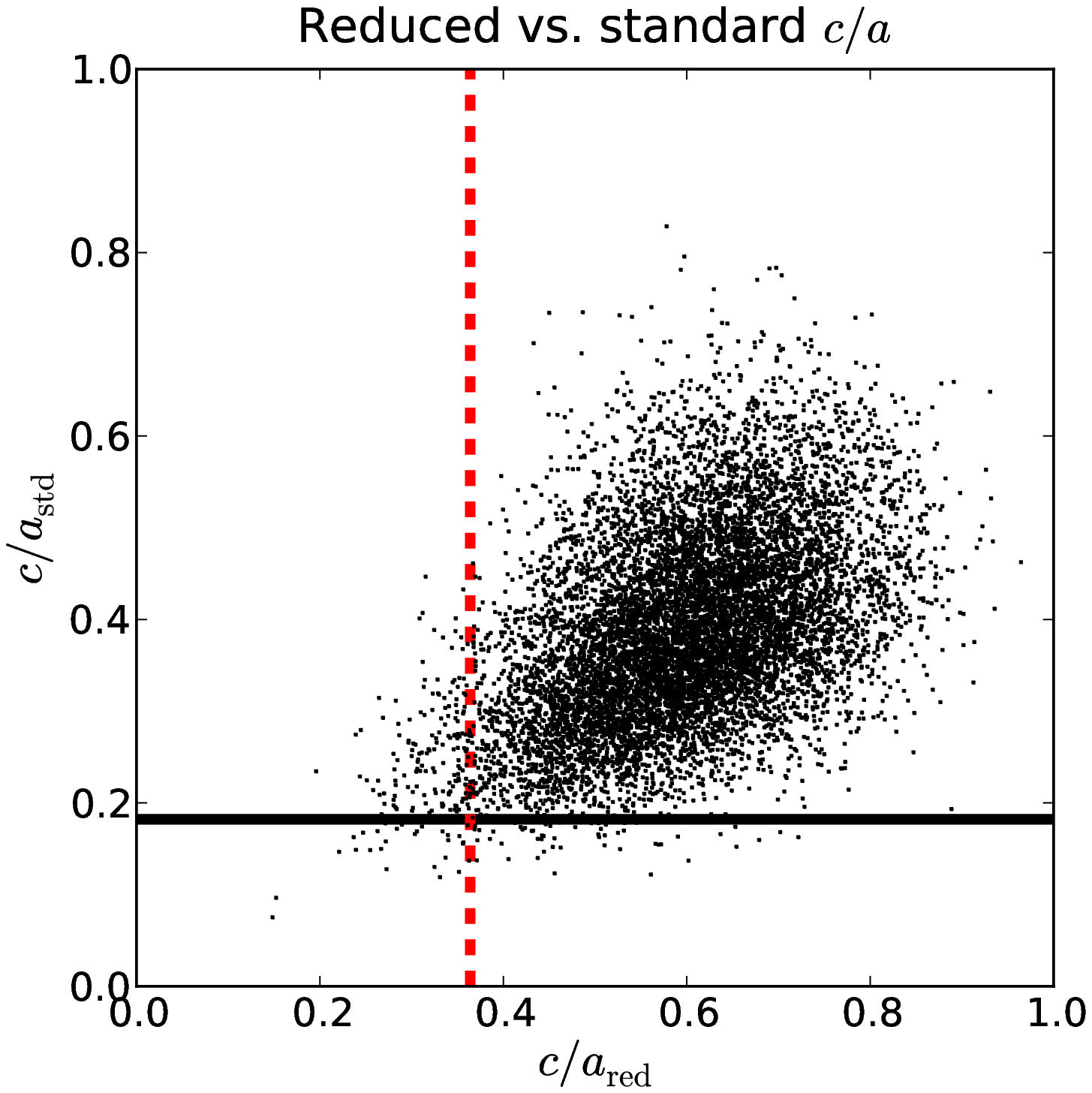}
   \includegraphics[width=80mm]{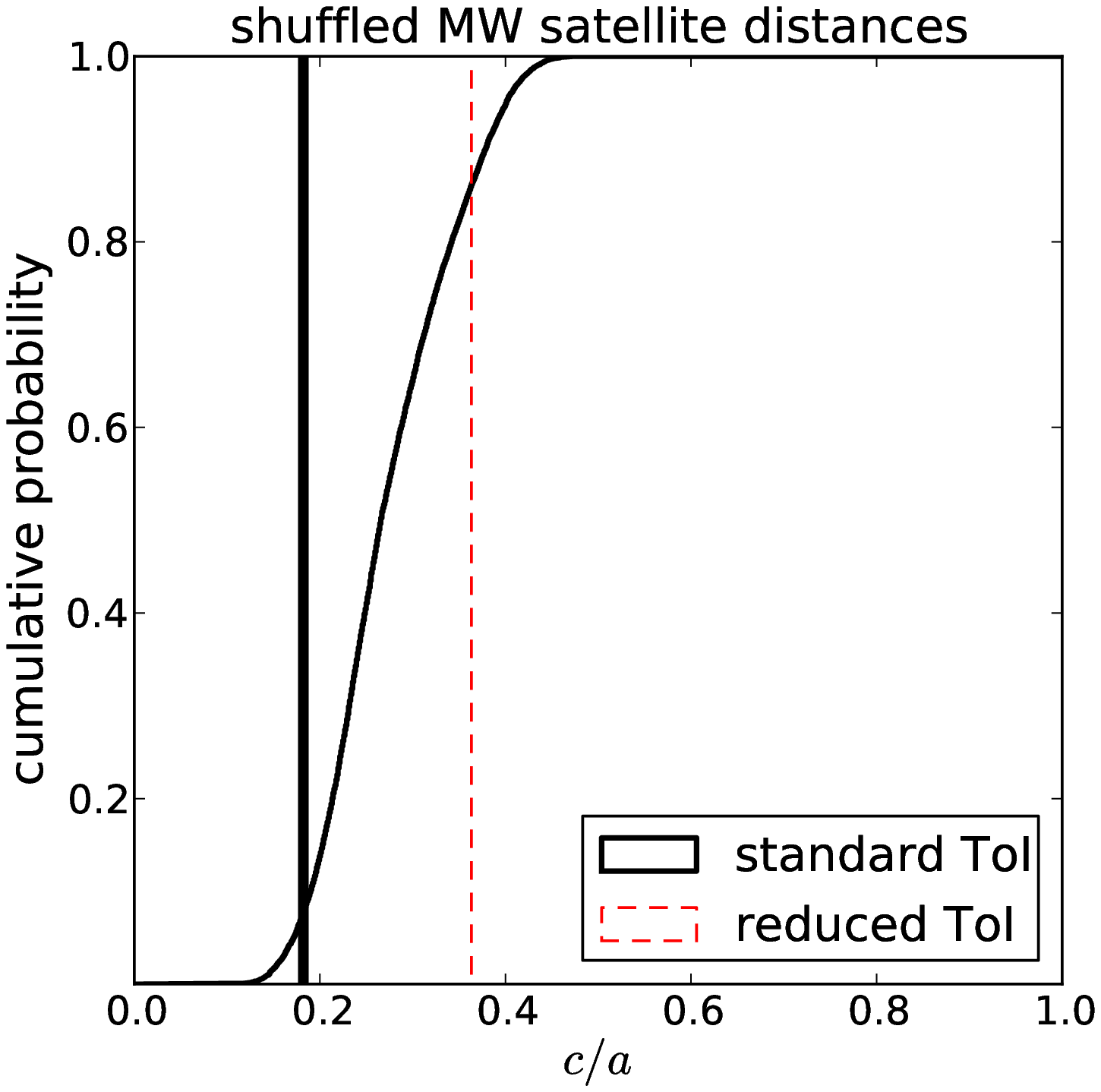}
   \includegraphics[width=80mm]{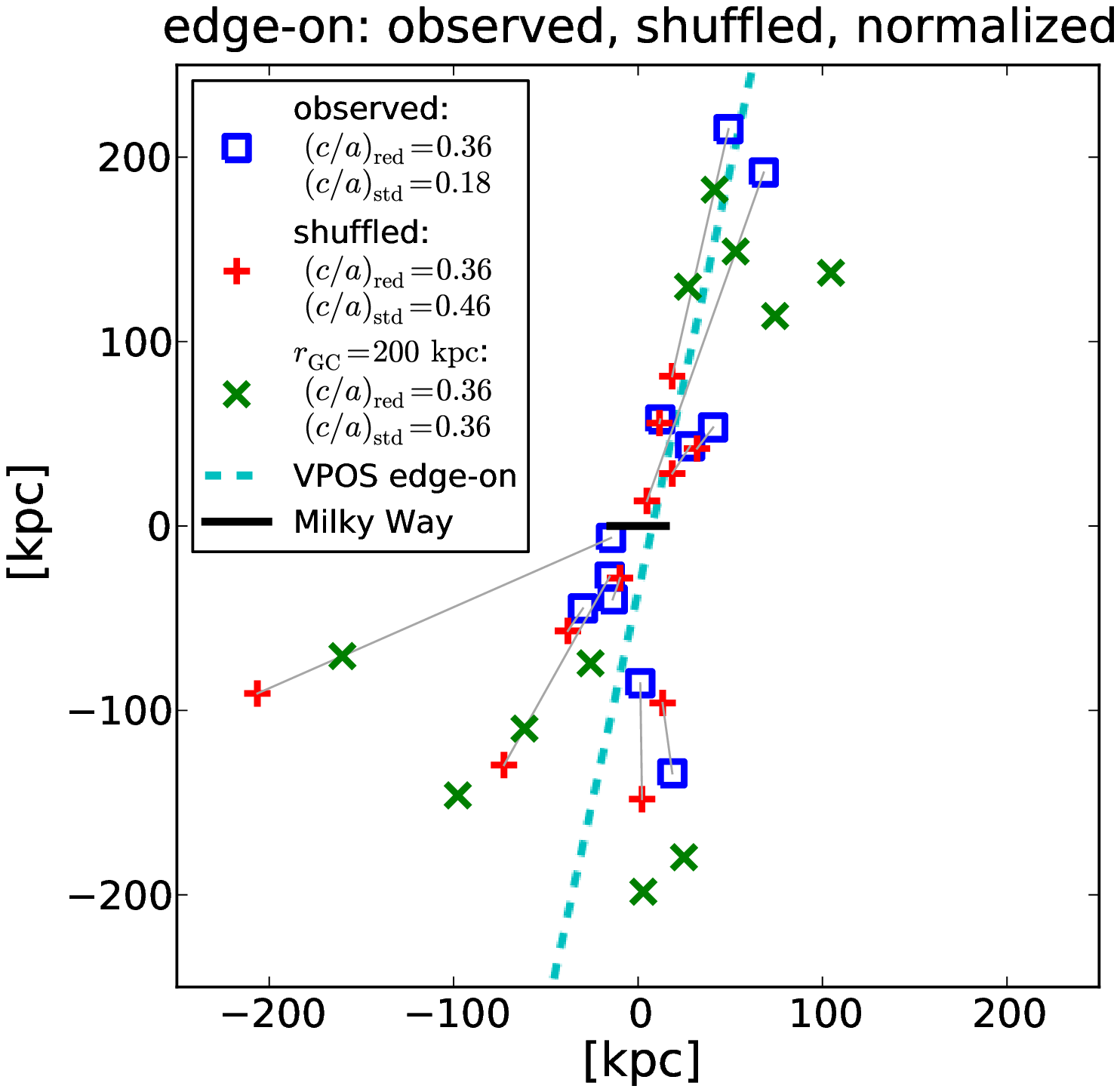}
   \caption{
   Comparison of the flattening of satellite systems determined with the reduced and with the standard ToI.
\textit{Upper left panel}: Cumulative distribution of $c/a$\ axis ratios of isotropic satellite distributions following the exact same radial distribution as the observed MW satellite galaxies. The axis ratios of the same set of 10000 random realisations are determined using the standard (black solid, $(c/a)_{\mathrm{std}}$) and reduced (red dashed, $(c/a)_{\mathrm{red}}$) ToI. The corresponding values for the 11 classical MW satellites, $(c/a)_{\mathrm{std}}^{\mathrm{MW}} = 0.18$\ and $(c/a)_{\mathrm{red}}^{\mathrm{MW}}$ = 0.36, are indicated with the thicker vertical lines. About one per cent of the random realisations reproduce the observed flattening measured via $(c/a)_{\mathrm{std}}$, whereas twice as many of the exact same realisations have a similarly small or smaller $(c/a)_{\mathrm{red}}$\ than the MW satellite system.
\textit{Upper right panel}: Comparison of axis ratios $(c/a)_{\mathrm{std}}$\ (y-axis) and $(c/a)_{\mathrm{red}}$\ (x-axis) for the same set of 10000 randomized satellite distributions. 
\textit{Lower left panel}: Cumulative distribution of $c/a$\ axis ratios similar to those in Fig. \ref{fig:positions}, but for 10000 satellite distributions constructed by keeping the observed angular positions fixed and randomly shuffling the observed satellite distances (without repetition). By definition, $(c/a)_{\mathrm{red}}$\ is the same for all such realizations because it is obtained from the reduced ToI, which ignored the radial distribution of the satellites. In contrast, the full three-dimensional ToI results in a range of axis ratios $(c/a)_{\mathrm{std}}$, most of which are larger than the value deduced from the observed MW satellites (solid, vertical black line). This indicates that the observed MW satellites are indeed flattened in three-dimensional space and not only in their positions on the sky.
\textit{Lower right panel}: The positions of the 11 classical MW satellites (blue squares) in Cartesian coordinates, seen along Galactic longitude $l = 247^\circ (= 157^\circ + 90^\circ)$, such that a plane fitted to the satellite positions is seen edge-on (cyan dashed line). The MW is situated at the origin, its disk is oriented edge-on as illustrated by the black line in the center. Satellites at larger radial distances tend to have a smaller angular distance from the best-fit plane.
The red plus signs illustrate one realization with shuffled satellite distances: both the angular distribution of the satellites as well as the radial distribution are preserved, but the distance-shuffled realisation is clearly less spatially flattened and has a much larger $(c/a)_{\mathrm{std}}$\ than the observed satellite distribution. However, a test which measures $(c/a)_{\mathrm{red}}$\ (using the reduced ToI) considers both situations to be equivalent. The same is true if all satellites are normalized to have a Galactocentric distance of $r_{\mathrm{GC}} = 200$\,kpc (green crosses). This illustrates that the full three-dimensional ToI method needs to be employed to test whether simulated satellite systems have spatial flattenings that are comparable to the MW satellite system.
   }
              \label{fig:positions}
\end{figure*}

To assess whether testing the flattening of a satellite distribution using the reduced ToI gives comparable results to using the standard ToI, we construct 10000 randomized satellite systems drawn from an isotropic distribution. One random realization is constructed by assigning a random angular position on a sphere (centered on the Galactic center) to each of the 11 classical MW satellite galaxies, while keeping each satellite's Galactocentric distance fixed. This ensures that the satellite distribution exactly follows the observed radial distribution of the MW satellites.
We then measure $(c/a)_{\mathrm{std}}$ and $(c/a)_{\mathrm{red}}$\ for each realization and determine whether they fall below the observed value for the MW satellite system. The resulting cumulative distributions are plotted in the upper left panel of Figure~\ref{fig:positions}.

We find that $(c/a)_{\mathrm{std}} \leq (c/a)_{\mathrm{std}}^{\mathrm{MW}}$\ in 1.05 per cent of all cases. The spatial flattening of the MW system measured via the minor-to-major axis ratio of the full three-dimensional positions of the 11 brightest satellites is thus very unlikely under the hypothesis that the satellite positions
 are isotropically distributed. For the reduced ToI method, using normalized satellite distances, we find $(c/a)_{\mathrm{red}} \leq (c/a)_{\mathrm{red}}^{\mathrm{MW}}$\ in 1.95 per cent of all cases. 
 Testing anisotropy using this method is roughly twice as likely to agree with an isotropic distribution. 
 In this sense, the reduced ToI test is less discriminating.

Of the 195 realizations fulfilling the $(c/a)_{\mathrm{red}}$\ criterion, only 38 simultaneously fulfill the $(c/a)_{\mathrm{std}}$\ criterion. The \textit{majority}, 81 per cent, of the realizations that pass the $(c/a)_{\mathrm{red}}$\ test are less flattened than the MW satellite system; that is: they have a larger axis ratio than observed. The upper right panel of Figure~\ref{fig:positions} demonstrates this point. The figure compares the axis ratios $(c/a)_{\mathrm{std}}$\ and $(c/a)_{\mathrm{red}}$\ for the set of 10000 randomized satellite distributions. While a weak correlation between the two measures is present, there is considerable scatter, and most realizations which have as low a $(c/a)_{\mathrm{red}}$\ as the MW satellite system (left of the red dashed line) are not simultaneously as flattened in their full 3D distribution measured via $(c/a)_{\mathrm{std}}$\ (below the black solid line).

We conclude that using the reduced ToI test as a statistic for the degree of spatial flattening produces 
more than 80 per cent false positives when applied to the MW satellite system. 
\textit{The reduced ToI test is an insufficient criterion for the comparison with the observed three-dimensional satellite distribution.}

\subsubsection{Fixed satellite positions but reshuffled distances}

The spatial flattening of a satellite systems implies a certain degree of correspondence between a satellite's distance, and its angular coordinate. The reduced ToI axis ratio $(c/a)_{\mathrm{red}}$\ is completely insensitive to this correspondence since it  considers only the normalized satellite positions. To demonstrate this, we generated 10000 ``reshuffled" satellite distributions, keeping the angular positions of the 11 classical MW satellites fixed but selecting their distances randomly, without replacement, from the observed set of 11 distances. This procedure again conserves the observed radial distribution since each distance is selected exactly once per realization.

The lower left panel of Fig.~\ref{fig:positions} shows the results. Not surprisingly, most of distance-reshuffled realizations (more than 92.5 per cent\footnote{We have checked that this number increases to 94.7 per cent (i.e. an even more pronounced planar anisotropy) if instead of the Galactocentric positions we use the Heliocentric positions and shuffle the Heliocentric distances of the observed MW satellites before they are transformed to a Galactocentric coordinate system for the ToI analysis.}) have larger $(c/a)_{\mathrm{std}}$\ than the observed MW satellite population and are therefore less strongly flattened. The MW satellite distribution is thus indeed plane-like in three-dimensional space and not merely anisotropic in the angular positions around the MW (in which case one would expect the axis ratios of shuffled realizations to more evenly distribute below and above the observed $(c/a)_{\mathrm{std}}^{\mathrm{MW}}$). The reduced ToI test, however, is by definition unable to discriminate between these different cases and results in the same $(c/a)_{\mathrm{red}}$ independent of the reshuffling.

The lower right panel of Fig.~\ref{fig:positions} provides a more intuitive illustration of this. It shows the positions of the 11 classical MW satellites (blue squares) such that the best-fit plane (determined using the \textit{reduced} ToI) is seen edge-on. In addition, the red plus signs mark the satellite positions in one of the reshuffled distributions: they clearly do not look like a flattened planar distribution. This impression is confirmed by their (standard ToI) axis ratio of $(c/a)_{\mathrm{std}} = 0.46$. However, by construction the reduced ToI axis ratio is the same as the observed one. Similarly, the green crosses mark the satellite positions if they would all reside at the exact same Galactocentric distance of 200\,kpc (i.e. their distances are normalized to 200\,kpc). Again their $(c/a)_{\mathrm{std}}$\ axis ratio is much larger than that of the observed satellite positions, but the reduced ToI test would classify this system to be just as flattened as the observed MW satellite system.

\textit{As expected, using the reduced ToI is not a robust test of the spatial flattening of satellite galaxy systems.} Ignoring the radial distances of the satellites when testing their spatial anisotropy biases towards finding agreement with the observed situation even though the full three-dimensional distribution can be markedly different.


\subsection{Kinematics}
\label{sect:velocities}

Here we suggest a simple, first-order test, to check whether an orbital pole distribution, for instance such as that of the EAGLE model satellites shown in fig. 4 of S14, indicates a similarly unexpected clustering of orbital poles close to the normal direction of the best-fitting plane to the satellite positions as is present for the observed data.

\subsubsection{The observed orbital poles}
\label{sect:poles}

\begin{figure*}
   \centering
   \includegraphics[width=140mm]{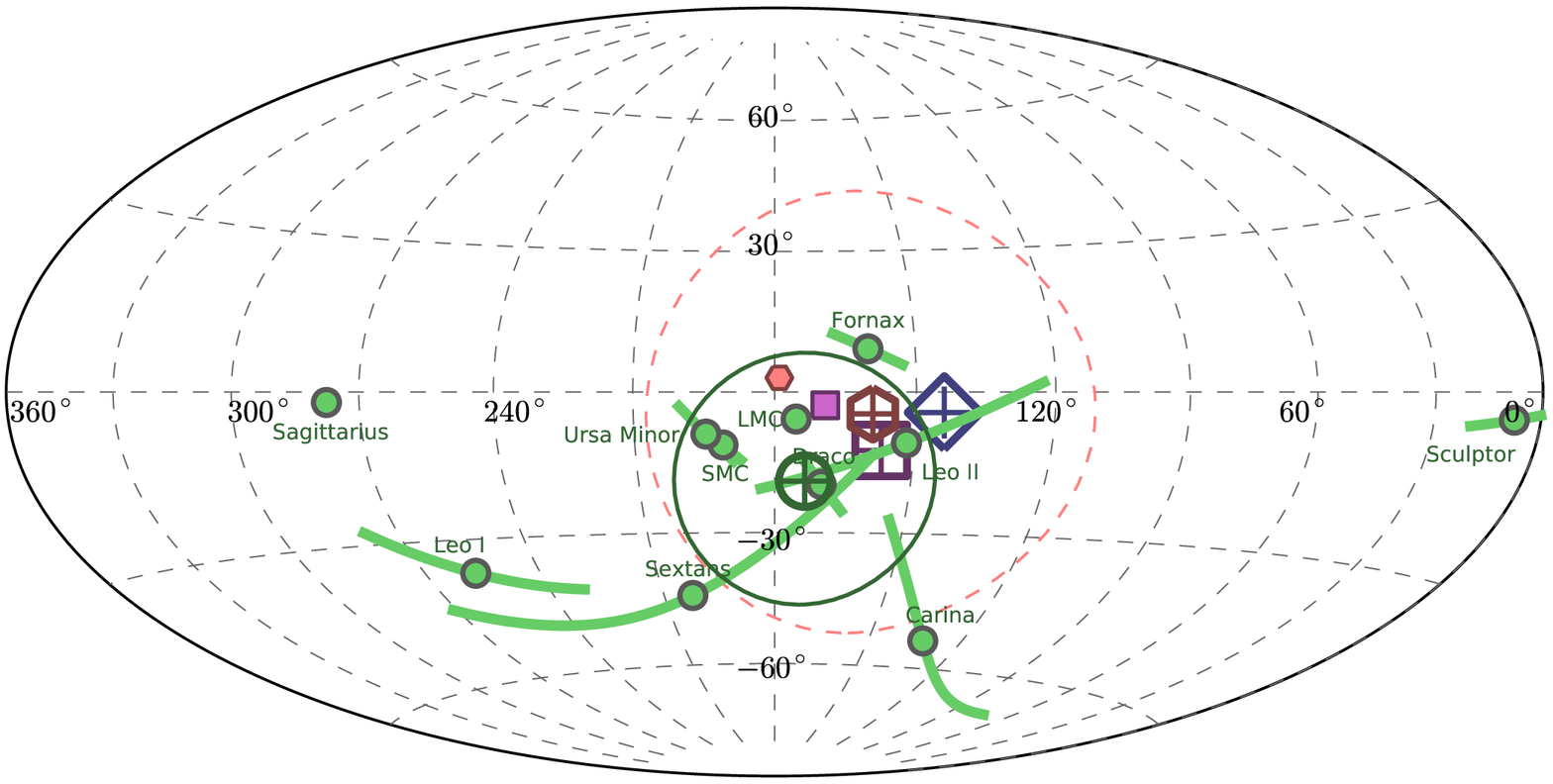}
   \includegraphics[width=140mm]{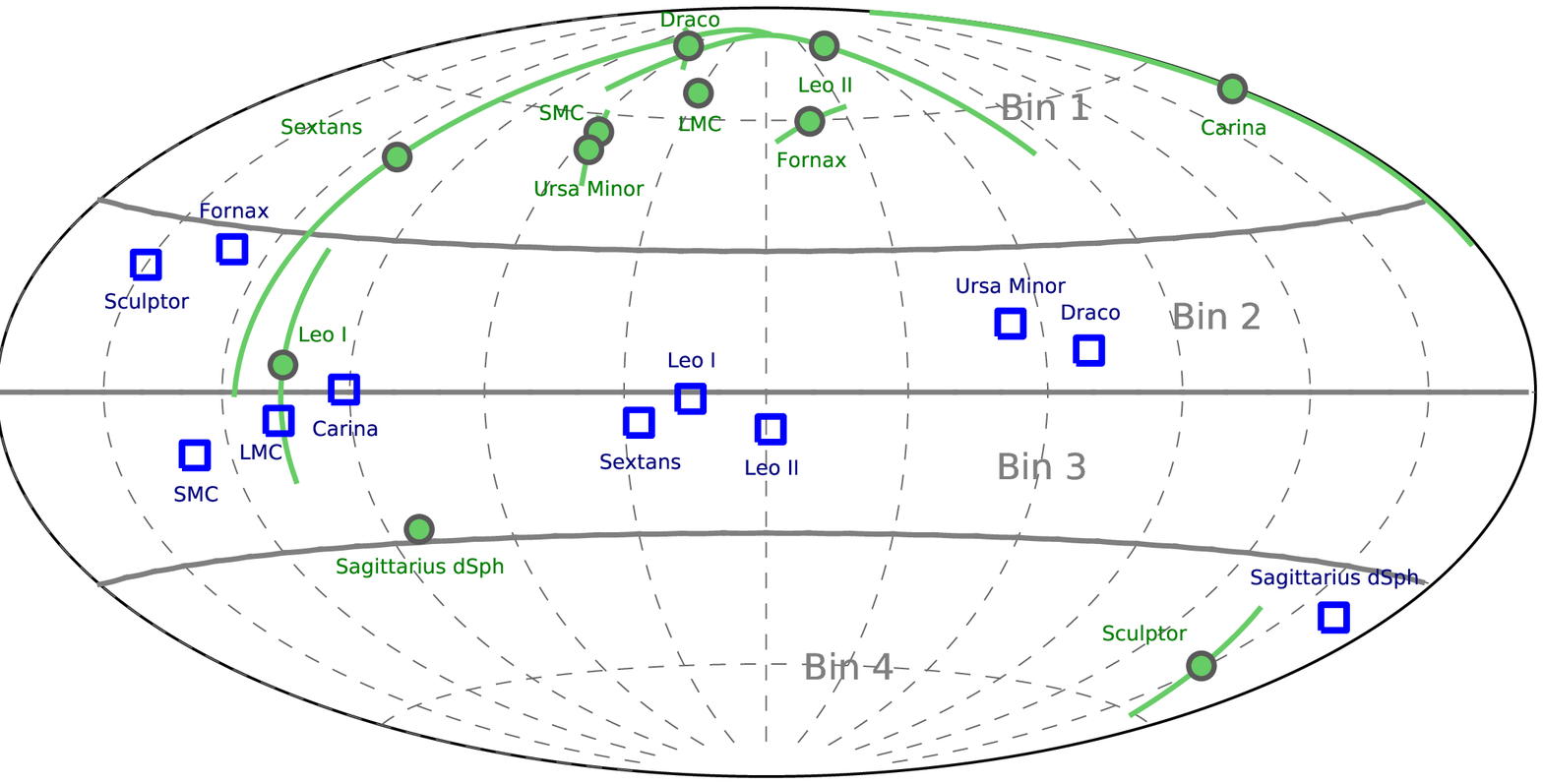}
   \caption{
\textit{Upper panel}: The orbital poles of the 11 brightest ('classical') MW satellites for which PMs are measured (green dots with green great-circle segments indicating the 1$\sigma$-uncertainties) are plotted in Galactic coordinates. 
Also shown are the normal to the best-fitting plane of the 11 classical MW satellites (light magenta square), the VPOS normal (dark magenta open square with plus sign), the normal to the plane fitted to the MW globular clusters classified as young halo objects (blue open diamond with plus sign), the average orbital pole direction (dark green open circle with plus sign, using the most-recent determination from an updated list of streams in \citealt{PawlowskiKroupa2014}), the average direction of all MW stream normals (dark red hexagon with plus sign, from \citealt{PawlowskiKroupa2014}), and the stream normal of the Magellanic Stream (small light-red hexagon). 
\textit{Lower panel}: The same orbital poles are plotted in a coordinate system defined by setting the 'north' and 'south' poles of the coordinate system to align with the minor axis determined using the reduced ToI method. 
The positions of the MW satellites (blue squares) are plotted in addition to their orbital poles (green dots). One can clearly see that most orbital poles cluster close to the 'northern' pole of the coordinate system, and thus close to the normal direction of the best-fitting plane. 
}
              \label{fig:poles}
\end{figure*}

For the MW satellites, we determine their orbital poles from their observed positions and proper motions as discussed in detail in \citet{PawlowskiKroupa2013}. If more than one proper motion measurement is available for a particular satellite, we use the uncertainty-weighted proper motion to determine the satellite's most-likely orbital pole direction. For Draco, we update the Hubble Space Telescope proper motion measurement from the preliminary value used in \citet{PawlowskiKroupa2013} to the one published in \citet{Pryor2014}. Compared to the orbital pole published in \citet{PawlowskiKroupa2013} this improves the alignment of Draco's orbital pole with the VPOS plane normal (angle to the normal of the best-fitting plane to the 11 classical satellites $\theta^{\mathrm{class}}_{\mathrm{VPOS}} = {14.1^{\circ}}_{-0.1}^{+2.6}$) and results in an even tighter concentration of orbital poles. The eight most-concentrated poles now have a spherical standard distance of $\Delta_{\mathrm{sph}} = 27.2^\circ$\ instead of $29.3^\circ$. \textit{This is in line with the trend that better proper motion data results in closer alignments of the derived orbital poles with the satellite plane axis and a reduction of the scatter around it.}

Fig. \ref{fig:poles} plots these observed orbital poles. The upper panel shows the orbital poles and the normal directions to the VPOS, the plane fitted to the young halo globular clusters \citep{Pawlowski2012a} and the average stream normal \citep{PawlowskiKroupa2014} in Galactic coordinates. The lower panel shows the poles in a coordinate system in which the 'north' and 'south' poles of the all-sky-plot are pointing along the minor axis of the distribution of the 11 classical MW satellites, as determined using the reduced ToI method. This makes the projected satellite positions, plotted as blue squares, to lie close to the 'equator' of this coordinate system.

\subsubsection{Binned orbital poles}
\label{Sect:binnedpoles}

\begin{figure}
   \centering
   \includegraphics[width=80mm]{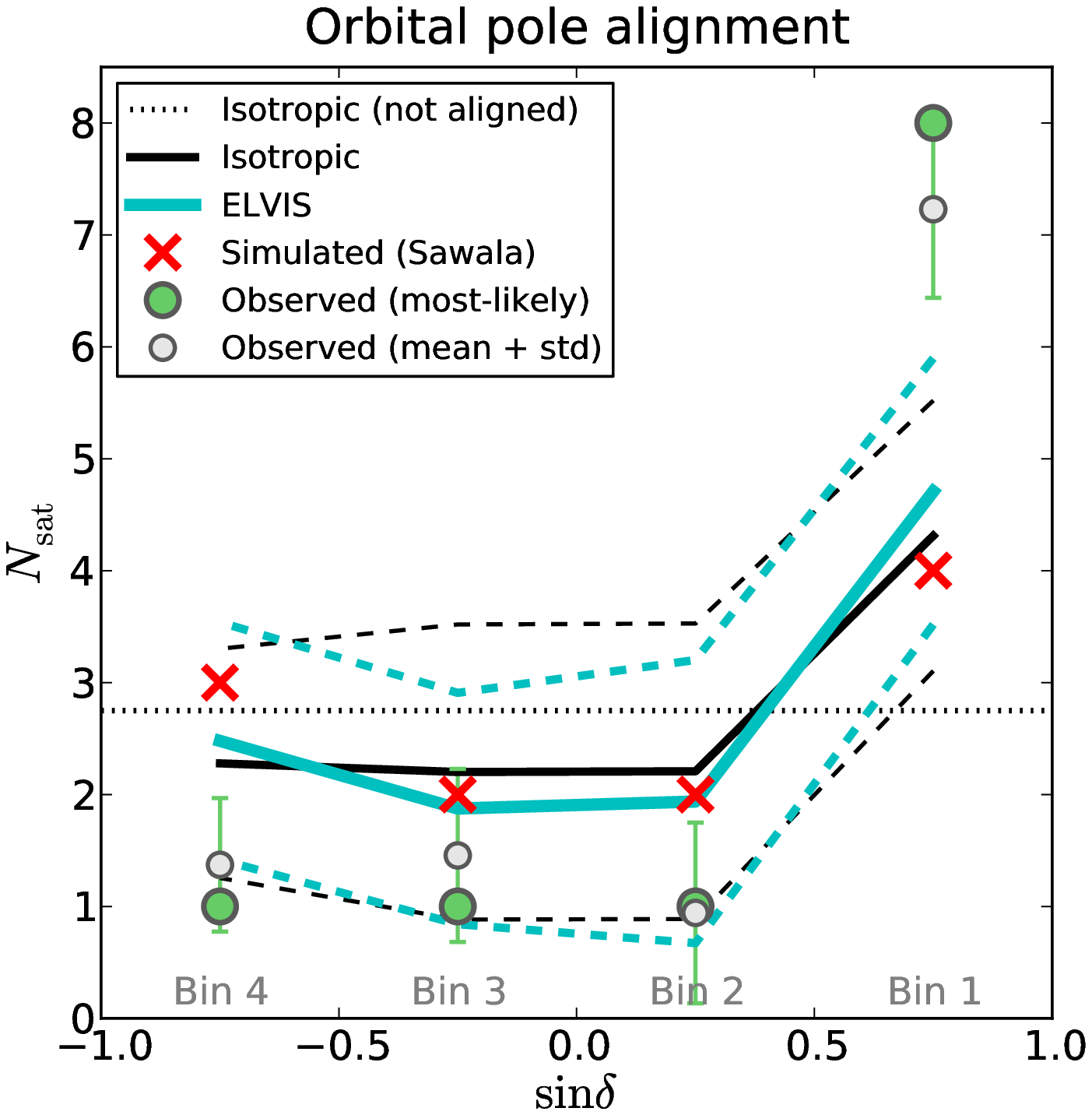}
   \caption{
Distribution of observed and expected orbital poles of 11-satellite systems among the four equal-area bins defined in Fig. \ref{fig:poles}. The large green dots illustrate the counts for the most-likely observed orbital poles, the open dots illustrate the mean counts obtained from 10000 realisations varying the observed proper motions within their uncertainties (compare to the orbital pole direction uncertainties represented by the green great-circle segments in Fig. \ref{fig:poles}). The error bars show the standard deviation around those means. The solid black line gives the average number of orbital poles per bin for isotropic satellite systems with isotropic velocity directions, the thicker solid cyan line gives the average number over the 48 halos in the ELVIS suite of collision-less cosmological simulations. Dashed lines indicate the $1\sigma$ scatter around the averages. The red crosses indicate the number of orbital poles per bin for the hydrodynamic simulation plotted in figure 4 of S14. They agree well with the average expected numbers. The lack of a strong orbital pole peak comparable to that of the observed poles in bin 1 reveals that the simulation does not contain a coherently rotating satellite plane comparable to that around the MW.
   }
              \label{fig:polebins}
\end{figure}

To test whether a simulated satellite system has an orbital pole distribution that is as strongly clustered as the observed orbital poles of the MW satellites, we start by drawing 100000 realizations of 11 satellite positions from an isotropic distribution with normalized distances. For each satellite, a velocity direction is selected at random from an isotropic distribution. Then the orbital pole of each of the satellites is constructed. 

Using the reduced ToI method, we determine the minor axis of the satellite distribution for each realization of 11 satellites. We then transform the satellite positions and orbital pole directions of a given realization into a 'satellite plane coordinate system': we align the minor axis with the poles of the coordinate system. Because the minor axis is an axial direction, we are left with two possibilities how to determine the 'northern' (upper) coordinate system pole. We decide to define it as that of the plane normal directions which has the larger number of orbital poles aligned to within $60^{\circ}$\ (one quarter of the sky). If both directions have the same number of aligned poles, one is chosen at random.

Eight of the observed MW satellites have orbital poles pointing to within $60^{\circ}$ of the satellite plane normal. Out of the 100000 random realizations, only 0.79 per cent have at least this observed number of 8 orbital poles within $60^\circ$\ (bin 1) of the northern plane pole. Even finding only at least 7 poles in this area is rare and occurs in 4.1 per cent of the realizations. This demonstrates that the observed orbital pole distribution for the MW is rare and significant at the 99 per cent level, but note that we are using a simplified test which does not utilize the full information available for the MW satellite orbital pole distribution such that this underestimates the significance \citep[for a better test see][]{PawlowskiKroupa2013}.

We define four equal-area bins in the satellite coordinate system. Bin 1 is defined as $< 60^\circ$ from the 'northern' pole, bin 2 as $\leq 30^\circ$\ 'north' of the equator, bin 3 as $\leq 30^\circ$\ 'south' of the equator and bin 4 covers the area $< 60^\circ$ from the 'southern' pole. They contain 8, 1, 1 and 1 of the observed MW satellite orbital poles, respectively. This allows one to easily compare the observed orbital pole distribution with those of simulations. As an example, we use the simulation results shown in fig. 4 of S14. From that plot, we read 4, 2, 2, and 3 orbital poles in these bins.

A perfectly isotropic orbital pole distribution would yield on average $\frac{11}{4} = 2.75$\ poles per bin (black dotted line in Fig. \ref{fig:polebins}). However, the coordinate system is defined by the angular positions of the satellites, which are by definition $90^\circ$\ away from the associated orbital pole (the direction of angular momentum around the origin is always perpendicular to the position vector). Thus, in a coordinate system such as ours which is defined to have an equator $90^\circ$\ away from the reduced ToI minor axis, a completely isotropic distribution of positions and velocities (which we sampled with the 100000 realizations discussed before) results in slightly more orbital poles close to the coordinate system's 'north' and 'south' poles, and thus bins 1 and 4, than would be the case for a completely randomly oriented coordinate system. The resulting distribution is plotted as the black solid line in Fig. \ref{fig:polebins}, the dashed lines indicate the standard deviation from this average (about $\pm 1.2$\ poles per bin).

The green dots in the figure illustrate the distribution of the most-likely positions of the observed orbital poles \citep{PawlowskiKroupa2013}. The strong peak of 8 poles within the 'northernmost' (right) bin 1 differs by more than three standard deviations from the expected number of about four in the case of an isotropic distribution. In contrast to the observed orbital poles, the simulated orbital poles (red crosses, read from fig. 4 of S14), follow the expectation derived from the isotropic distributions of satellite velocities if the effect of defining the coordinate system by the satellite positions on the sky is taken into account. For this example, our test thus concludes that there is no indication that the simulation of S14 reproduces a similarly strongly rotationally supported disk of satellites like the one around the MW.

To estimate how much the proper motion uncertainties of the observed MW satellites affect this comparison, we have generated 10000 realisation of orbital pole distributions by randomly varying the proper motions within their uncertainties. For each realisation the counts in bins 1 to 4 were recorded. Fig. \ref{fig:polebins} shows the resulting average counts (smaller open dots) and the standard deviations around them (error bars). Even though the orbital poles of Sextans and Carina have large uncertainties which can move them out of bin 1, on average there are more than 7 orbital poles in bin 1 and that number does not drop below 6. Thus, even though some of the observed orbital poles have considerable uncertainties, the tension with S14 remains significant when accounting for them.

An extreme kinematic coherence is the most important defining characteristic of the VPOS (and also the GPoA and other satellite galaxy planes). A simulation which does not contain a similar number of closely concentrated orbital poles as the observed MW satellite system can therefore not be claimed to resolve the disk of satellites problem. In particular, it is clear that the simulation analyzed and plotted by S14 does not have as strong a correlation of orbital poles with the normal direction to the satellite plane as the observed MW satellite system.


\subsection{Comparison to collision-less simulations}
\label{sect:dmonly}

\begin{figure}
   \centering
   \includegraphics[width=80mm]{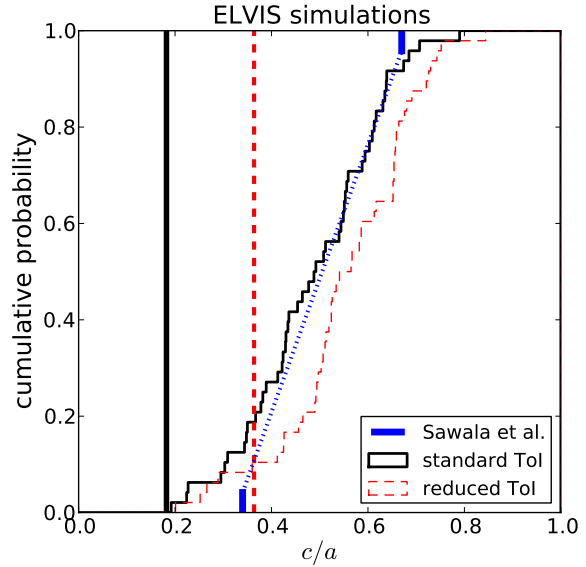}
   \caption{
Cumulative distribution of $(c/a)_{\mathrm{red}}$\ and $(c/a)_{\mathrm{std}}$\ axis ratios as in Fig. \ref{fig:positions}, but for satellites in the ELVIS suite of dark-matter-only simulations of Local Group analogs \citep{GarrisonKimmel2014a}. The blue lines indicate the range of $(c/a)_{\mathrm{red}} = 0.34\ \mathrm{to}\ 0.67$\ reported by S14 for the hydrodynamical simulations including baryonic physics, determined using the reduced ToI. We do not know the shape of the cumulative distribution between these points. 
   }
              \label{fig:elvis}
\end{figure}

Satellite positions and velocities are expected to be largely independent of baryonic physics, in particular on the scale of hundreds of kpc which is relevant for the disk of satellites problem. However, S14 claim to have resolved the 'plane of satellites' problem (among others) by simultaneously simulating baryonic and dark matter in Local Group equivalents,

The EAGLE simulation results are not publicly available, but we are nevertheless able to test the dark-matter-only case by analyzing a different set of cosmological zoom simulations of Local Group analogs. The 'Exploring the Local Volume in Simulations' (ELVIS) suite of simulations \citep{GarrisonKimmel2014a} consists of 12 pairs of main halos with masses, separations, and relative velocities similar to those of the MW and M31 and a control sample of 24 isolated halos matching the paired ones in mass. The simulations are dissipationless, they do not model the effects of baryons. The results of these simulations, in particular the final $z = 0$\ sub-halo positions and velocities, are publicly available\footnote{http://localgroup.ps.uci.edu/elvis/}. The ELVIS suite has already been used to demonstrated that satellite systems around paired hosts are not more likely to contain VPOS-like planes, and that VPOS-like structures are extremely rare: only one out of 4800 realizations displayed a satellite system that is similarly flattened and has similarly concentrated orbital poles as the 11 classical MW satellites \citep{PawlowskiMcGaugh2014b}. That study followed the standard tests of root-mean-square plane height and $(c/a)_{\mathrm{std}}$\ axis ratios determined via the standard ToI using the full three-dimensional positions, and also tested the concentration of orbital poles.

In the following, we analyze the satellite systems of the 48 ELVIS host halos in an equivalent way to S14's analysis of the satellite systems around the 24 EAGLE hosts using the $(c/a)_{\mathrm{red}}$\ axis ratio. We follow the standard abundance-matching assumption that the 11 most-luminous satellites are embedded in those 11 sub-halos which had the largest peak-mass during their lifetime. If the addition of baryonic physics to the simulation indeed results in more anisotropic or flattened satellite distributions using this measure, the ELVIS satellite systems should tend to have larger $(c/a)_{\mathrm{red}}$\ than the range of 0.34 to 0.67 reported by S14.

Fig. \ref{fig:elvis} compiles our results. We find that the range of $(c/a)_{\mathrm{red}}$\ reported for the EAGLE hydrodynamical simulations is well contained within the range of $(c/a)_{\mathrm{red}} = 0.20$\ to 0.84 found for the ELVIS simulations. Choosing the 11 sub-halos which are most-likely to contain the most-luminous satellite galaxies in a dark-matter-only simulation not only results in the same range of $(c/a)_{\mathrm{red}}$, but in fact in a larger range than that reported for the hydrodynamical simulation. This is not entirely unexpected, because the ELVIS simulation suite contains twice as many host halos. If we restrict our analysis to the 20 hosts in a paired, Local Group like configuration (excluding two host pairs where a third nearby massive halo is present), we find $(c/a)_{\mathrm{red}} = 0.25$\ to 0.74, a more narrow range but still exceeding that of the hydrodynamic simulations in both directions.

Nine of the 48 ELVIS satellite systems have a larger $(c/a)_{\mathrm{red}}$\ than 0.67, the upper limit of the range reported by S14. Five of the ELVIS main halos host satellite systems have reduced ToI axis ratios below $(c/a)_{\mathrm{red}}^{\mathrm{MW}} = 0.36$. Four of these also have $(c/a)_{\mathrm{red}}$\ below the lowest value found by S14 in their hydrodynamical simulations. \textit{Thus, there is no indication that the addition of baryonic physics to cosmological simulations of Local Group equivalents results in more flattened satellite distributions.} In analogy to Figure \ref{fig:positions}, we have also determined the full three-dimensional axis ratio $(c/a)_{\mathrm{std}}$\ using the standard ToI method. The resulting cumulative distribution is shown as a black line in Figure \ref{fig:elvis}. The observed $(c/a)_{\mathrm{std}}^{\mathrm{MW}} = 0.18$\ is not reproduced by any of the 48 ELVIS satellite systems, again illustrating that using the reduced ToI is not sufficient to test whether a simulated satellite system is as flattened as the observed MW satellite population. A flattening similar to that of the classical satellite galaxies of the MW, as measured using the reduced ToI method, is not a unique feature of hydrodynamic cosmological simulations. 

We have also calculated the distribution of orbital poles for each of the ELVIS satellite systems in the same coordinate system as used in Sect. \ref{sect:velocities}. The result is part of Figure \ref{fig:polebins}: the cyan solid line indicates the average number of orbital poles per bin found from using the 11 brightest satellites from each of the 48 ELVIS simulations. The dashed line again indicates the standard deviation in each bin. We find that the average number of ELVIS orbital poles for the different bins closely follows the trend as present for the isotropic satellite positions and velocities. The ELVIS orbital poles are a little more aligned with the satellite plane normal as illustrated by the slightly higher signal in bins 1 and 4. This qualitatively agrees with earlier studies that found that sub-halo based satellite systems are similar to -- but slightly more anisotropic than -- satellite distributions drawn from isotropy \citep{Metz2007,Pawlowski2012b,PawlowskiMcGaugh2014b}. In this binned analysis, the orbital pole distribution found in the EAGLE hydrodynamical simulations is fully consistent with the average numbers of orbital poles extracted from the dark-matter-only simulation.

\textit{Thus we have to conclude that there currently is no evidence which supports claims that accounting for baryonic physics in cosmological simulations improves the chances to find co-orbiting planes of satellite galaxies as pronounced as observed around the MW.} Non-dissipative dark-matter-only simulations give comparable results, and do {\it not} solve the problem, as has long been well known.


\section{Too-big-to-fail Problem}
\label{sect:TBTF}

\begin{figure}
   \centering
   \includegraphics[width=80mm]{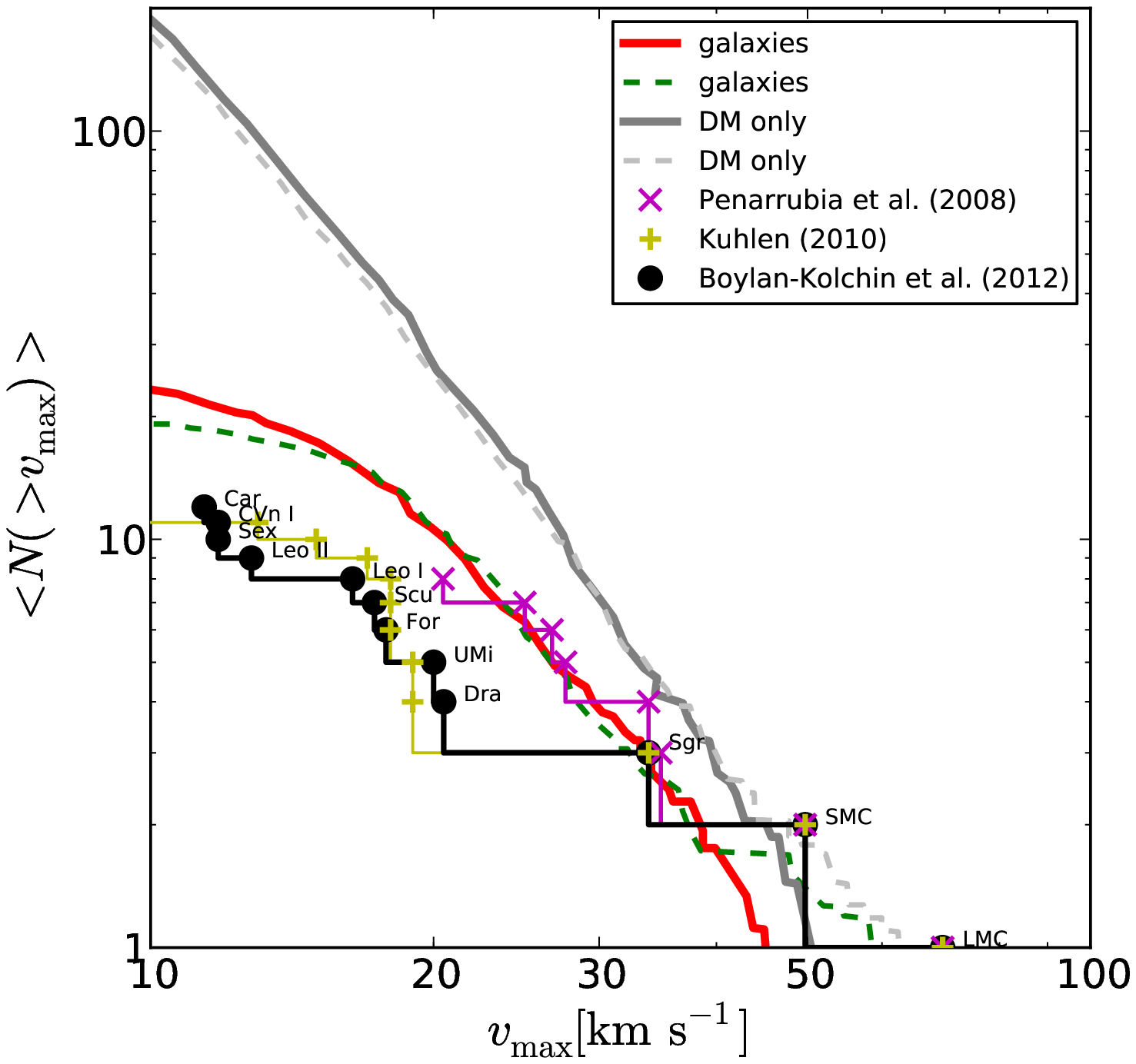}
   \caption{
Cumulative number of satellites (sub-halos or galaxies) as a function of their maximum circular velocity $v_{\mathrm{max}}$. The data for the EAGLE simulations as presented in S14 were extracted from their figure 3. These are averages of the 12 simulations of Local Group equivalents analyzed by S14. The two upper (grey) lines show the cumulative sub-halo number for their dark-matter-only simulations, the two lower (red and green) lines the cumulative number of luminous satellite galaxies for the MW and M31 equivalents.
The magenta crosses are the old $v_{\mathrm{max}}$\ estimates for the observed MW satellites by \citet{Penarrubia2008}. S14 compared only these with their simulated galaxies. They conclude that their simulations have resolved the TBTF problem because these MW satellite data points follow the curve of simulated satellite galaxies (but see Sect. \ref{sect:TBTF} for why this is an inaccurate definition and test of the TBTF problem). The yellow plus signs and black dots are the $v_{\mathrm{max}}$\ of the observed MW satellites by \citet{Kuhlen2010} and \citet{BoylanKolchin2012}, respectively. The latter are labelled with the satellite names. In particular the more carefully determined velocities of \citet{BoylanKolchin2012} demonstrate that the TBTF problem is still present. For velocities below $v_{\mathrm{max}} \lesssim 25\,\mathrm{km\,s}^{-1}$, the hydrodynamical simulations contain about twice as many luminous galaxies than observed.
}
              \label{fig:TBTF}
\end{figure}

The TBTF problem can be formulated as follows: the measured densities at the (de-projected) half-light radii $r_{1/2}$\ of local galaxies are too low compared to the most massive subhalos predicted in $\Lambda$CDM simulations \citep{BoylanKolchin2012}. The enclosed dynamical mass (and thus density) within $r_{1/2}$, or its corresponding circular velocity $v_{\mathrm{circ}}$, can be directly and accurately estimated from the observed line-of-sight velocity dispersion \citep{Wolf2010}.

In contrast, some theoretical papers, such as S14, chose to compare the maximum circular velocity $v_{\mathrm{max}}$\ of galaxies formed in their simulation to that of the MW satellites. This velocity can not be directly measured for observed dwarf galaxies. Inferring $v_{\mathrm{max}}$\ from observational data requires assumptions about how the dark matter density profile extrapolates from $r_{1/2} \approx 500\,\mathrm{pc}$, where the kinematics are measured, to the radius $r_{max} \approx 1~\mathrm{to}~5\,\mathrm{kpc}$, where the circular velocity curve of the satellite's expected dark matter halo has its maximum. Assigning $v_{\mathrm{max}}$\ values to observed satellites is therefore very uncertain and depends on the assumptions made. 
Consequently, comparing simulations to observations based on $v_{\mathrm{max}}$\ is considerably less straightforward than the comparison of $v_{\mathrm{circ}}$\ at $r_{1/2}$, which for best accuracy can use the simulation particle data directly \citep[see e.g.][]{BoylanKolchin2012,Tollerud2014}. Due to these possible systematics, counting satellites above a given $v_{\mathrm{max}}$\ obscures the underlying issue of the TBTF problem and does not directly address it. We therefore refer to this formulation as the \textit{extrapolated TBTF problem}.

A meaningful comparison with the observed MW satellites requires a careful assessment of the best available observational information, in particular when the method of comparison can be severely affected by systematics. S14 only compare their simulated galaxies with the $v_{\mathrm{max}}$\ values of \citet{Penarrubia2008}. To estimate $v_{\mathrm{max}}$\ for the observed MW satellites, that study assumes that the dark matter halos follow a NFW profile \citep{Navarro1996}, and makes use of the relation between $v_{\mathrm{max}}$\ and $r_{\mathrm{max}}$\ as given by field halos in $\Lambda$CDM. Thus, already the comparison of these velocities to the dark-matter-only simulations by S14 is problematic (the analysis is concerned with satellite sub-halos, not with isolated field halos), but even more so for the hydrodynamical simulations which result in less concentrated sub-halos (having a lower central dark matter density) than dark-matter-only simulations \citep{Zolotov2012}. Furthermore, more recent estimates result in significantly lower values of $v_{\mathrm{max}}$\ for the same MW satellites \citep{Kuhlen2010,BoylanKolchin2012}.

To demonstrate the differences in the estimated $v_{\mathrm{max}}$, we plot the cumulative distribution of the number of sub-halos above a given $v_{\mathrm{max}}$\ in Figure \ref{fig:TBTF}. The curves for the simulated galaxies (and sub-halos for the dark-matter-only runs) have been extracted from figure 3 of S14 by carefully tracing them. In addition to the $v_{\mathrm{max}}$\ values \citet{Penarrubia2008} report for the MW satellites (magenta crosses), we also plot the two more recent estimates (yellow plus signs and black dots). In particular the $v_{\mathrm{max}}$\ values by \citet{BoylanKolchin2012} are significantly lower than those of \citet{Penarrubia2008}. According to \citet{BoylanKolchin2012}, this difference is due to their much more detailed analysis. They compare the observed properties of the MW satellites with a large number of sub-halos from the Aquarius simulations \citep{Springel2008}, use more recent dynamical constraints based on the mass within the half-light radius (which has smaller uncertainties than previous mass estimates, \citealt{Wolf2010}), and even correct for the effect of the gravitational softening length on the inner halo densities. Comparing the resulting $v_{\mathrm{max}}$\ values of \citet{BoylanKolchin2012} instead of those of \citet{Penarrubia2008} with the simulations of S14 shows that the hydrodynamical simulations merely alleviated but do not resolve the extrapolated TBTF problem. Below $v_{\mathrm{max}} \lesssim 25 \mathrm{km\,s}^{-1}$, the simulations contain about twice as many satellites with a given velocity than observed.

Fig. \ref{fig:TBTF} illustrates that the value of $v_{\mathrm{max}}$\ that is assigned to a satellite depends very sensitively on theoretical priors. It is not a direct measurement. None of these issues would arise if, to test whether the TBTF problem is resolved, the observed $v_{\mathrm{circ}} (r_{\mathrm{1/2}})$\ would have been directly compared with the rotation curves of the simulated galaxies \citep[as for example done by][]{BoylanKolchin2012,Tollerud2014}. However, even at the highest resolution level L1, S14 still have a softening length of $\epsilon = 94$\,pc, and for their extrapolated TBTF problem comparison they use resolution level L2, which has $\epsilon = 216$\,pc.
According to \citet{BoylanKolchin2012}, softening has the effect to reduce the density on scales of about $3\epsilon$, such that cumulative quantities like $v_{\mathrm{circ}}$\ are underestimated. The de-projected half-light radii of several of the classical MW satellites are of the same order, five out of the nine dSphs analyzed in \citet{BoylanKolchin2012} have $r_{1/2}$\ between 200 and 400\,pc. It therefore appears unlikely that the simulations of S14 reliably resolve the radii relevant for the TBTF problem. Their might even lack the resolution to conclusively investigate the problem at all.

Finally, Figure \ref{fig:TBTF} (or figure 3 of S14) also shows that there appears to be no problem at the high-mass end of the cumulative $v_{\mathrm{max}}$\ distribution for the dark-matter-only simulations (grey lines). If using the \citet{Penarrubia2008} velocities, the dark-matter-only distribution matches the satellite data above $30\,\mathrm{km\,s}^{-1}$. The dark-matter-only simulations begin to over-predict the $v_{\mathrm{max}}$\ function only at lower masses, therefore it looks as if the extrapolated TBTF problem was solved without the need for any hydrodynamics. One likely reason for the low number of simulated halos above $v_{\mathrm{max}} = 30\,\mathrm{km\,s}^{-1}$\ is that the host halos are at the low-mass end of what is reasonable for the MW and M31. Already \citet{BoylanKolchin2012} have discussed that a significantly less massive MW halo could in principle avoid the TBTF problem, but their estimate of the required halo mass of about $0.5 \times 10^{12}$\ appears unrealistically small (see the discussion in their section 5.1, but also \citealt{Wang2012,VeraCiro2013}). According to the supplementary information of S14, the second-most-massive galaxies in their Local Group equivalent simulations (which can be identified as the MW equivalents) have a median mass of only $0.9 \times 10^{12}$\ and a lower limit of $0.5 \times 10^{12}$. This means that half of the MW equivalents have halo masses between 0.5 and $0.9 \times 10^{12}$.

Similarly, the stellar masses of the MW and M31 analogs in the S14 hydrodynamical simulations lie in the range of 1.5 to $5.5 \times 10^{10}\,\mathrm{M}_{\sun}$. These are \textit{very} low compared to the observed values. The lower limit of this range, $1.5 \times 10^{10}\,\mathrm{M}_{\sun}$, is extremely low compared to the stellar mass of the MW ($4.9~\mathrm{to}~5.5~\times~10^{10}\,\mathrm{M}_{\sun}$; \citealt{Flynn2006}) and even the upper end of the range, $5.5 \times 10^{10}\,\mathrm{M}_{\sun}$, is very low compared to the stellar mass of M31 ($10~\mathrm{to}~15~\times~10^{10}\,\mathrm{M}_{\sun}$; \citealt{Tamm2012}). This again suggests that the associated halos are of too low mass, resulting in fewer sub-halos for a given $v_{\mathrm{max}}$\ and thus alleviating the TBTF problem before any hydrodynamical effects come into play.


\section{Summary}
\label{sect:conclusion}

The addition of baryonic physics to $\Lambda$CDM models of hierarchical structure formation can ameliorate
certain discrepancies between the observed and simulated structures of galaxies.
In this paper, we investigate recent claims 
that baryonic physics can do more,  providing a solution even to the ``satellite planes'' problem, 
the nearly planar distribution of satellite galaxies around the Milky Way and other galaxies.
Such claims are surprising given that the scale of the satellite systems (hundreds of kiloparsecs)
is much greater than the scale over which baryonic physics can be expected to act.
By analyzing dissipationless simulations in a consistent way to a previous analysis of dissipational simulations,  
we find no significant difference between the predicted distribution of satellite plane flattenings.
Claims to the contrary were shown to result in part from a non-standard measure of plane
thickness that ignores the radial positions of the satellites.

In addition to being thin, the Milky Way satellite system also exhibits coherent rotation.
We introduced a simple test for kinematic coherence that compares the alignment of orbital poles
with the satellite plane normal. 
We apply the new test to the results of a hydrodynamical simulation and show that the latter
fail to reproduce the concentration of orbital poles observed in the Milky Way satellite system. 
We find that the distribution of orbital poles  in the hydrodynamical simulation
is consistent with that produced by random, isotropic velocities, as in the dark-matter-only simulations.
Here again, we find no evidence that the addition of baryonic physics is useful in reconciling 
 simulations with observations.

Yet another issue of $\Lambda$CDM is the ``too-big-to-fail'' (TBTF) problem.
TBTF refers to the high densities of the inner parts of simulated sub-halos compared with the central
densities of observed satellites; the latter computed using the circular velocity, $v_{1/2}$, 
measured at the half-light radius $r_{1/2}$. 
Recent claims that the TBTF problem is resolved in hydrodynamical simulations are based on a comparison 
of the {\it maximum} circular velocities, $v_{\mathrm{max}}$, of simulated and observed galaxies.
Since $v_\mathrm{max}$ can not be measured for the Milky Way satellites,
this method relies on a model-dependent extrapolation of the mass distribution.
When we account for the fact that the simulated satellites are situated in sub-halos, 
not field halos, we find that the TBTF problem is still present, even in its extrapolated form.
Once again, confrontation of simulations with observational data shows no preference for
hydrodynamical over dark-matter-only simulations.


\acknowledgments
We thank Shea Garrison-Kimmel and the ELVIS collaboration for making their simulations publicly available. 
The contribution of MSP to this publication was made possible through the support of a grant from the John Templeton Foundation. DM was supported by the National Science Foundation under grant no. AST 1211602 and by the National Aeronautics and Space Administration under grant no. NNX13AG92G.

\end{document}